\documentclass[11pt]{article}
\usepackage{amsmath,amstext,amsbsy,amssymb}
\usepackage{bm}
\newcommand{\fr}{\frac}
\newcommand{\lb}{\label}

\newcommand{\prm}{\prime}

\newcommand{\be}{\begin{equation}}
\newcommand{\ee}{\end{equation}}
\newcommand{\ba}{\begin{array}}
\newcommand{\ea}{\end{array}}
\newcommand{\beqa}{\begin{eqnarray}}
\newcommand{\eeqa}{\end{eqnarray}}

\newcommand{\al}{\alpha}
\newcommand{\bt}{\beta}
\newcommand{\ka}{\kappa}
\newcommand{\la}{\lambda}

\newcommand{\si}{\sigma}

\newcommand{\te}{\theta}

\newcommand{\ga}{\gamma}

\newcommand{\om}{\omega}

\newcommand{\dgr}{\dagger}
\newcommand{\del}{\partial}

\newcommand{\nno}{\nonumber}

\newcommand*{\bgc}{\begin{center}}
\newcommand*{\edc}{\end{center}}
\newcommand*{\bgt}{\begin{table}}
\newcommand*{\edt}{\end{table}}

\def \A {{\mathcal A}}

\newcommand{\h}{\hat}

\long\def\symbolfootnote[#1]#2{\begingroup%
\def\thefootnote{\fnsymbol{footnote}}\footnote[#1]{#2}\endgroup}

\begin{document}

\begin{center}

{\Large \bf An Alternative Formulation of Hall Effect 
and Quantum Phases in Noncommutative Space}

\vspace{0.5cm}

{\bf{\"{O}mer F. Day\i}$^{a,b,}$
\symbolfootnote[0]{{\it E-mail addresses:} dayi@itu.edu.tr and
dayi@gursey.gov.tr; barisy@gursey.gov.tr }} \,{and}\, {\bf{Bar\i \c{s}
Yap\i\c{s}kan}$^{c}$ }\\
\vspace{5mm}

{\em $^{a}${\it Physics Department, Faculty of Science and
Letters, Istanbul Technical University,\\
TR-34469, Maslak--Istanbul, Turkey} }

{\em $^{b}${\it Feza G\"{u}rsey Institute, P.O. Box 6, TR-34684,
\c{C}engelk\"{o}y, Istanbul, Turkey } }

{\em $^{c}${\it Physics Department, Faculty of Science and
Letters, Mimar Sinan Fine Arts University,\\
\c{C}\i ra\u{g}an Cad. \c{C}i\u{g}dem Sok. No:1,  TR-34349, Be\c{s}ikta\c{s}
--Istanbul, Turkey}}

\vspace{2cm}

{\bf Abstract}

\end{center}

A recent method of constructing quantum mechanics in noncommutative coordinates, 
alternative to implying noncommutativity by means of star product  is
discussed. Within this approach we  study Hall effect  as well 
as quantum phases in noncommutative coordinates. The $\theta$--deformed  phases  which
we obtain are  velocity independent.

\newpage

\section{Introduction}
Formulation of quantum mechanics avoiding operators  was given at the beginning of quantum era by Weyl--Wigner--Groenewold--Moyal (WWGM)\cite{wwgm}
as the $\hbar$--deformation of classical mechanics:
Observables are functions taking values in the classical phase space $(P_I,Q_I)$ whose combinations are constructed through the star product
\begin{equation}
\label{SCT}
\star   =
\exp \left[
\frac{i\hbar }{2} \left(
\frac{\overleftarrow{\del}}{\del Q^I}
\frac{\overrightarrow{\del}}{\del P_I}
-\frac{\overleftarrow{\del}}{\del P_I}
\frac{\overrightarrow{\del}}{\del Q^I}
\right) \right] ,
\end{equation}
where $\overleftarrow{\del}$ $(\overrightarrow{\del})$ indicates
that the derivative is applied to the left (right). Although the WWGM
formulation of quantum mechanics has some shortcomings, like the
lack of positive definite probabilities, it turned out to be crucial
to establish quantum mechanics in noncommutative coordinates: One
treats coordinates as commutative and imply
the noncommutativity by the star product given with an
antisymmetric, constant deformation parameter $\theta_{IJ}$  as
\begin{equation}
\label{NSP}
\star_\theta   \equiv
\exp \left[
\frac{i}{2}\theta_{IJ}
\frac{\overleftarrow{\del}}{\del Q^I}
\frac{\overrightarrow{\del}}{\del Q^J}
 \right] .
\end{equation}
$\theta$--deformed  Hamiltonian  systems are built inserting the star product (\ref{NSP})  between the bilinear or higher terms
appearing in the original quantum Hamiltonians. Indeed, this is equivalent to the shift of coordinates
\begin{equation}
\label{SH}
 Q_I\rightarrow  Q_I-\frac{1}{2\hbar}\theta_{IJ}  P^{op}_J,
\end{equation}
where $P_I^{op} =-i\hbar\frac{\del}{\del Q^I}$  is used. As it is obvious when the original Hamiltonian does not possess any coordinate
dependence this procedure will not furnish any $\theta$--deformation. Especially, it is not adequate to consider spin degrees of freedom.

There is an alternative method of defining
quantum Hamiltonians in noncommutative spaces\cite{oj-09}
as far as in the starting Hamiltonian there exist terms which can be interpreted as minimally coupled gauge fields
which may be either functions or matrix valued.
In this work we would like to discuss this alternative procedure of obtaining dynamical systems in noncommuting coordinates
and apply it to some interesting physical systems where the Hamiltonians are functions or matrices.

Employing the star product (\ref{NSP}) or the equivalent shift (\ref{SH}), diverse physical systems
like Hall effect in noncommuting coordinates\cite{ao} and quantum phases in noncommutative spaces\cite{ao,glr,cpst,mz,wl,prfn} were considered.
In all these dynamical problems there are some external fields which can be interpreted as gauge fields
interacting with the particles in terms of the minimal coupling procedure.
Within the alternative method  we first
show that Hall effect in noncommutative coordinates can be formulated where resulting Hall conductivity is $\theta$--deformed or not,
depending on the realization adopted. Then, we apply the new deformation procedure to obtain    velocity
independent formulations of
Aharonov--Bohm (AB)\cite{ab}, Aharonov--Casher (AC)\cite{ac}, He--McKellar--Wilkens (HMW)\cite{hmw} and Anandan\cite{anan,flr} phases
in noncommutative coordinates.
Most of the earlier formulations yielded velocity dependent quantum phases in noncommutative spaces,
in spite of the fact that the distinguished property of the original phases is their independence  from the velocity of  scattered particles.
We presented  a unified  formulation of the $\theta$--deformed quantum phases yielding velocity dependent terms.
The difference between them and our
method is clarified. In the last section we
discuss the results obtained and their consequences. In particularly we discuss how to select the suitable realization.

\section{The Alternative $\theta$--deformation of Quantum Mechanics}

Generating classical mechanics as the $\hbar \rightarrow 0$ limit  of  quantum mechanics can be best perceived by the WWGM method.
Let $(P_I , Q_I )$ denote the classical variables corresponding to
the quantum phase space variables $( P^{op}_I , Q^{op}_I);$
$I =1,\cdots ,M.$  Multiplication of observables in
the former  space is given by the usual operator product,
however in the latter  the star product (\ref{SCT}) is employed to carry out multiplications.
In the WWGM approach, to imitate quantum commutators one introduces the Moyal brackets
$$
[f(P,Q ) , g(P,Q ) ]_\star \equiv f(P,Q) \star g(P,Q ) - g(P,Q ) \star f(P,Q ) ,
$$
where  the observables $f(P,Q )$ and $g(P,Q )$  are some functions. Hence the
classical limit  of the commutators is equivalent to
\begin{equation}
{\lim_{\hbar \rightarrow 0}} \fr{-i}{\hbar } [f(P,Q ) , g(P,Q ) ]_\star
= \{ f(P,Q) , g(P,Q) \} \equiv
  \frac{\partial f}{\partial Q^I} \frac{\partial g}{\partial P_I}
-\frac{\partial f}{\partial P_I}\frac{\partial g}{\partial Q^I} ,\label{ocl}
\end{equation}
which is the Poisson bracket.
When the observables are  matrices  whose elements are  $M_{kl}(P,Q)$ and $N_{kl}(P,Q),$
the Moyal bracket is
\begin{equation}
\label{MM}
\left( [M(P,Q) , N(P,Q ) ]_\star \right)_{kl}  =  M_{km}(P,Q ) \star N_{ml}(P,Q )
- N_{km}(P,Q ) \star M_{ml}(P,Q ) .
\end{equation}
However, its classical limit (\ref{ocl}),
in addition to the Poisson brackets, yields
a singular commutator of matrices.
Hence, instead of the classical limit (\ref{ocl}) we
deal with the  ``semiclassical'' limit defined by retaining the  terms
up to  $\hbar $ of  the Moyal bracket (\ref{MM}):
\begin{equation}
 \{M(P,Q ) , N(P,Q ) \}_C \equiv
\frac{-i}{\hbar}[M,N]+
\frac{1}{2} \{ M(P,Q ) , N(P,Q) \}
-\frac{1}{2} \{ N(P,Q ) , M(P,Q ) \} . \label{SCB}
\end{equation}
The first term is the
commutator of matrices, it should not be confused with the quantum mechanical commutator.
Obviously, the bracket (\ref{SCB}) does not satisfy Jacobi identities. However, we consider a semiclassical approach
in which the semiclassical limit  is taken after performing multiplication of the observables in terms of the star product.
Hence, vanishing of  the first two terms of the Moyal bracket relation
$$
\frac{-i}{\hbar}\left(\{M,\{ N,L \}_\star \}_\star +
\{N,\{ L,M \}_\star\}_\star  +
\{L,\{ M,N \}_\star\}_\star \right) = {\cal O}_{-1} (\frac{1}{\hbar})+{\cal O}_0 (\hbar^0)+{\cal O}_1 (\hbar) +\cdots
$$
should be considered in  the semiclassical limit of the Jacobi identity.
Indeed, one can show that the semiclassical limit of the Jacobi identity
\beqa
{\cal O}_{-1} (\frac{1}{\hbar})+{\cal O}_0 (\hbar^0) & = &
-\fr{i}{\hbar}[M,[N,L]]+[M, \{ N , L \}]
-[M , \{ L , N\}] \nonumber \\
&&
+\{M ,[N,L]\}
 -\{[N,L],M\}
 + ({\rm cyclic\ permutations\ of}\  M,N,L) =0 \nonumber
\eeqa
is satisfied.
Moreover, one can observe that the semiclassical limit of the Leibniz rule defined  as
$$
\{M\star N, L\}_C=\{M,L\}_C\star N+ M\star \{N,L\}_C ,
$$
is also satisfied at the $\hbar$ order.
Although, these considerations are essential for a consistent definition of the semiclassical method of matrix observables, in this work we will
deal with a quantum phase space algebra obtained somehow utilizing the bracket (\ref{SCB}). Obviously, adopted operator realization
of commutators among the phase space variables should  satisfy the usual Jacobi identities as we will discuss.

Dynamical systems in noncommutative space can be formulated in terms of
the  following first order matrix Lagrangian (for the most general case see \cite{o1}),
by choosing the coordinates as $Q_I=(r_\al ,p_\al),$ where
 $\al,\beta =1,\cdots ,{\rm d}, $
\begin{equation}
\label{lag}
L   = {\dot r}^\al \left[ \frac{p_\al}{2}\mathbb{I} +\rho A_\al (r) \right]
-\frac{{\dot p}^\al}{2} \mathbb{I}\left[r_\al  + \frac{\theta_{\al \beta}}{ \hbar} p^\beta  \right] - H_0(r,p) .
\end{equation}
$\mathbb{I}$ denotes the unit matrix possessing the same dimension with the matrix valued gauge field $\A_\al .$ We denoted the coupling constant as  $\rho$.
Although (\ref{lag}) is classical, $\hbar$ is present to furnish the constant, antisymmetric noncommutativity parameter $\theta_{\al \beta }$ with
the dimension $\left(\sf{length}\right)^2$.
The canonical momenta  $P_I=(P_r^\al =\del L/ \del \dot r_\al ,\  P_p^\al =\del L / \del \dot p_\al),$
corresponding to the coordinates $Q_I=(r_\al ,p_\al),$
yield the dynamical  constraints
\begin{eqnarray}
\psi^{1\al} & \equiv & (P_r^\al -\fr{1}{2}p^\al )\mathbb{I} -\rho \A^\al ,\nonumber \\ 
\psi^{2\al} & \equiv & (P_p^\al +\fr{1}{2}r^\al )\mathbb{I} +\frac{\theta^{\al \beta}}{ \hbar} p_\beta . \nonumber 
\end{eqnarray}
They satisfy the semiclassical relations
\begin{eqnarray}
\{\psi^1_\al ,\psi^1_\beta \}_C & = & \rho F_{\al \beta} ,\nonumber \\
\{\psi^2_\al ,\psi^2_\beta \}_C & = & \frac{\theta_{\al \beta}}{ \hbar},\nonumber \\
\{\psi^1_\al ,\psi^2_\beta \}_C & = & -\delta_{\al \beta} , \nonumber
\end{eqnarray}
where we introduced the field strength:
\begin{equation}
F_{\al \beta} =
\fr{\partial A_\beta}{\partial r^\al}
-\fr{\partial A_\al}{\partial r^\beta}
-\fr{i\rho }{\hbar} [A_\al ,A_\beta ] .\label{fst}
\end{equation}
We would like to emphasize the fact that commutators appearing in this semiclassical formulation are the ordinary matrix commutators.
Obviously, $\psi^z_\al ;\ z=1,2,$ are second class constraints, so that one can take them into account by introducing
the semiclassical Dirac bracket defined as
\begin{equation}
\label{sdb}
\{M,N\}_{CD} \equiv \{M,N\}_C -\{M,\psi^z\}_C \ {\cal C}^{-1}_{zz^\prime}\ \{\psi^{z^\prime},N\}_C  ,
\end{equation}
in terms of ${\cal C}^{-1}$ which is
the inverse of
$$
{\cal C}^{zz^\prime}_{\al \beta}=
\{\psi^z_\al , \psi^{z^\prime}_\beta\}_C .
$$
Therefore,
omitting $\mathbb{I}$ on the left hand sides, one can show that
the following relations are satisfied,
\begin{eqnarray}
\{r^\al,r^\beta \}_{CD}  & = &
 \frac{ \theta^{\al\beta  }}{\hbar} , \label{rr}\\
\{p^\al,p^\beta \}_{CD}  & = &
\rho  F^{\al\beta  }   -\frac{\rho^2}{\hbar}(F\te F)^{\al\beta} ,  \label{yy}\\
\{r^\al,p^\beta \}_{CD}  & = &
\delta^{\al \beta } - \frac{\rho}{\hbar} (\te F)^{\al\beta} ,   \label{ry}
\end{eqnarray}
keeping the terms at the first order in $\theta$ and  at the second
order in $\rho .$
We abbreviated $(\te F)^{\al\beta} \equiv \te^{\al\ga}F_{\ga}^{\beta} ,\ (F\te
F)^{\al\beta} \equiv F^{\al\ga} \te_{\ga}^\si F_{\si}^{\beta}$.

The semiclassical brackets (\ref{SCB})  differ from the Poisson brackets up to commutators of matrices, so that for observables which are
not matrices (\ref{sdb})  reduce to the ordinary Dirac brackets. Therefore, we can extend the canonical quantization rules to embrace the matrix observables
by the substitution of the basic brackets (\ref{rr})--(\ref{ry})
with the quantum commutators as
$\{, \}_{CD} \rightarrow \frac{1}{i\hbar} [ ,]_q .$ Then, we are furnished with
the generalized algebra
\begin{eqnarray}
{[\hat r^\al ,\hat r^\beta ]_q}  & = &
 i \theta^{\al\beta  } ,\label{rrq}\\
{[ \h p^\al,\h p^\beta ]_q } & = &
i\hbar \rho  F^{\al\beta  }   -i\rho^2(F\te F)^{\al\beta} , \label{yyq}\\
{[\h r^\al,\h p^\beta ]_q } & = &
i\hbar \delta^{\al \beta } - i\rho (\te F)^{\al\beta} ,  \label{ryq} \\
{[\h p^\al , \h r^\beta ]_q}  & = &
-i\hbar \delta^{\al \beta } + i\rho (F \te )^{\al\beta}   .\label{yrq}
\end{eqnarray}
We denoted quantum commutators by the subscript $q$ to distinguish them from matrix commutators.
Because of being  first order in $\theta$, the right hand sides of
(\ref{rrq})--(\ref{yrq}) may only possess ${\h r}_\al|_{\theta =0} =r_\al $ dependence.
Hence $F_{\al\beta}$ is still as in (\ref{fst}).
This is the starting point of the alternative method of establishing  quantum mechanics in noncommutative coordinates.
A realization of the generalized algebra (\ref{rrq})--(\ref{yrq}) and a Hamiltonian $H_0(p,r)$ should be provided.
Let us  deal with the operators
\beqa
\hat{p}_\al &=& D_\al
-\fr{\rho}{2\hbar}F_{\al\bt}\te^{\bt\ga}D_\ga, \label{R1}\\
\hat{r}_\al &=& r_\al -\fr{1}{2\hbar}\te_{\al\bt}D^\bt \label{R2},
\eeqa
where the covariant derivative is
$$
D_\al = -i\hbar\frac{\del}{\del r^\al}-\rho A_\al .
$$
One can demonstrate
that (\ref{R1}) and (\ref{R2})
satisfy the algebra  (\ref{rrq})--(\ref{yrq}) and the Jacobi identities, as far as the conditions
\be
\label{cons}
  -i\hbar\nabla_\al F_{\beta \ga}  -\rho [ A_\al , F_{\beta \ga}] = 0,\quad  [F_{\al \beta },F_{\ga \delta }]= 0
\ee
are fulfilled.
To illustrate the method let the initial Hamiltonian be $H_0 (p)=p^2/2m .$
Substituting $p$ with the quantum operator (\ref{R1}) one obtains the $\theta$--deformed Hamiltonian
\be
\label{IHAM}
H_0 (\h p)\equiv {\h H}_{nc}= \frac{1}{2m} \left( D_\al
-\fr{\rho}{2\hbar}F_{\al\bt}\te^{\bt\ga}D_\ga \right)^2 .
\ee
Setting $\theta =0$ yields the Hamiltonian operator
\be
\label{IH}
\h H= \frac{1}{2m} \left(  -i\hbar\frac{\del}{\del
r_\al}-\rho A_\al\right)^2 .
\ee
Therefore, (\ref{IHAM}) gives the noncommutative dynamics corresponding to the Hamiltonian  (\ref{IH}).

Let us present another representation of the algebra (\ref{rrq})--(\ref{yrq})  which will be utilized in the subsequent sections.
As far as (\ref{cons})  are valid, one can prove that
\beqa
\lb{rea21} \hat{p}_\al &=& -i\hbar\nabla_\al + \fr{\rho}{2}
F_{\al\bt}(r^\bt+ 2i \te^{\bt\ga}\nabla_\ga), \\
\lb{rea22} \hat{r}_\al &=& r_\al
-\fr{1}{2\hbar}\te_{\al\bt}(-i\hbar\nabla^\bt-\fr{\rho}{2}F^{\bt\ga}r_\ga)\eeqa
constitute  another realization of the algebra (\ref{rrq})--(\ref{yrq}). It is worth mentioning  that
in this representation only the gauge invariant field strength $F_{\al\beta}$ appears, in contrast
to (\ref{R1})--(\ref{R2}) where the gauge field $A_\al$ explicitly appears.

\section{Hall Effect in Noncommutative Coordinates}

We would like to study  the Hall effect in  noncommutative
space in the light of the alternative method prescribed in Section 2,
adopting different realizations.
This problem was addressed previously in \cite{ao} where the
Hall conductivity was shown to acquire a $\theta$--deformation factor.
However this resulted to the cost of  an unnatural overall factor in the definition of electric current.
We will show that
$\theta$--deformation of the Hall conductivity appears naturally in some realizations. Though in \cite{om}
a  natural $\theta$--deformed Hall conductivity was achieved, it was within the semiclassical approach of Section 2.

We consider
an electron moving on the two-dimensional plane $r_i=(x, y)$ in the presence  of the
uniform external in-plane electric field $E$ and the uniform external perpendicular
magnetic field $ B.$ The latter is taken into account by
the field strength
$F_{ij}=\epsilon_{ij}B,$
and fixing $\rho=-e/c .$
Hence the  generalized algebra (\ref{rrq})--(\ref{yrq}) yields
\begin{eqnarray}
{[\hat x ,\hat y ]_q}  & = & i \theta  ,\label{rrqH}\\
{[ \h p_i,\h p_j ]_q } & =  & -\frac{ieB\hbar}{c} \left(1 - \frac{eB\theta}{\hbar c}  \right)\epsilon_{ij}, \label{yyqH}\\
{[\h r_i,\h p_j ]_q } & = & i\hbar \left( 1 - \frac{e\theta B}{\hbar c} \right) \delta_{ij}  .\label{yrqH}
\end{eqnarray}
We choose the electric field to lie in the direction of the $x$--axis. Thus the Hamiltonian is taken  as
\be
\label{H0}
H = \fr{1}{2m}\hat{p}_i^2 + e E \hat{x} .
\ee
First  consider  the  realization of (\ref{rrqH})--(\ref{yrqH})
given in (\ref{R1})-(\ref{R2}) by choosing the symmetric gauge $A_i=\fr{eB}{2c}\epsilon_{ij}r_j,$
but  ignoring the terms at the
$e^2/c^2$ order:
\beqa
\hat{p}_i^{(1)} &=& (1-\fr{e\te B}{2\hbar c})p_i - \fr{eB}{2c}\epsilon_{ij}r^j ,\lb{rir1} \\
\hat{r}_i^{(1)} &=& (1-\fr{e\te B}{4\hbar c})r_i -\fr{\te}{2\hbar}
\epsilon_{ij}p^j . \lb{rir2}
\eeqa
By plugging  (\ref{rir1}) and (\ref{rir2}) into (\ref{H0})
one obtains the Hamiltonian
\be
\lb{hr11}
H^{(1)}=\fr{1}{2m}\left[ (1-2\ka)p_i-\fr{eB}{2c}\epsilon_{ij}r^j
\right]^2 +eE(1-\ka)x-\fr{eE\te}{2\hbar}p_y ,
\ee
where we defined
$p_i\equiv -i\hbar\nabla_i$ and $ \ka=\fr{e\te B}{4\hbar c}.$
On the other hand  keeping the terms at the $e^2/c^2$ order yields  the following realization
\beqa
\hat{p}_i^{(2)} &=& (1-\fr{e\te B}{2\hbar c})p_i -
\fr{eB}{2c}(1-\fr{e\te B}{2\hbar c})\epsilon_{ij}r^j ,\lb{rir3} \\
\hat{r}_i^{(2)} &=& (1-\fr{e\te B}{4\hbar c})r_i -\fr{\te}{2\hbar}
\epsilon_{ij}p^j .\lb{rir4}
\eeqa
When these are substituted in (\ref{H0}), one gets the $\theta$--deformed Hamiltonian as
\be \lb{hr12}
{H}^{(2)}=\fr{1}{2m}(1-2\ka)^2\left[ p_i -
\fr{eB}{2c}\epsilon_{ij}r^j \right]^2 + eE(1-\ka)x -
\fr{eE\te}{2\hbar}p_y .
 \ee

Another realization of (\ref{rrqH})--(\ref{yrqH}) can be read from
(\ref{rea21})--(\ref{rea22}): It does not refer to the explicit form of the vector field,
\beqa \hat{p}_i^{(3)} &=& (1-\fr{e\te B}{\hbar c})p_i
- \fr{eB}{2c}\epsilon_{ij}r^j ,\lb{rir5} \\
\hat{r}_i^{(3)} &=& (1+\fr{e\te B}{4\hbar c})r_i -\fr{\te}{2\hbar}
\epsilon_{ij}p^j . \lb{rir6}
\eeqa
Plugging (\ref{rir5}) and (\ref{rir6}) into (\ref{H0})
will produce the deformed Hamiltonian
\be \lb{hr2} {H}^{(3)} =
\fr{1}{2m}\left[(1-4\ka)p_i-\fr{eB}{2c}\epsilon_{ij}r^j \right]^2 +
eE(1+\ka)x - \fr{eE\te}{2\hbar}p_y .
\ee

It is not surprising that there exist different $\theta$-deformations of an underlying Hamiltonian. 
However, as we will argue in Section 5, preferring one to others is possible by adopting an
interpretation of the $\theta$-deformation and then specifying the adequate $\theta$-deformed physical quantities
derived  from the deformed Hamiltonians.   

Now we will examine these Hamiltonians in detail. In order to discuss the eigenvalue problem
of the Hamiltonians we perform the following change of variables
$$
{z}= {x}+i {y}, \qquad {p}_z = \fr{1}{2}({p}_x-i {p}_y) .
$$
Let us introduce two sets of creation and annihilation operators:
 \beqa
b = 2i \ga {p}_z + \fr{eB}{2c}\bt{\bar{z}} + \la_-; &
d = 2i\ga p_z - \fr{eB}{2c}\bt{\bar{z}}, &\nonumber \\
b^\dgr = -2i\ga {p}_{\bar{z}} + \fr{eB}{2c}\bt {z} + \la_-  ;&
d^\dgr = -2i\ga p_{\bar{z}}- \fr{eB}{2c}\bt{z} .& \nonumber
\eeqa
The constant coefficients $\ga,\ \bt$ and $\la_-$ will be fixed for each
Hamiltonian separately. These two mutually commuting sets  of operators
satisfy the commutation relations
$$
[b, b^\dgr] =
2m\hbar\ga\bt\om, \qquad\qquad [d, d^\dgr]= -2m\hbar\ga\bt\om ,
$$
where $\om=eB/mc$ is the cyclotron frequency. Each of the  Hamiltonians
(\ref{hr11}), (\ref{hr12}) and (\ref{hr2}) can be written in terms
of creation and annihilation operators   in the  form
\be
\label{HG}
{H} = \fr{1}{4m}(b b^\dgr + b^\dgr b) - \fr{\la_+}{2m}(d+ d^\dgr) - \fr{\la_-^2}{2m} ,
\ee
where the constant  $\la_+$ is also
going to be fixed for each Hamiltonian separately. The natural definition of the current operator
is
\be
 {{J}_i}=
 \fr{ie\rho_e}{\hbar}[ H,r_i ]= \fr{e\rho_e
 \ga}{m}(\ga{p}_i -\bt \fr{eB}{2c}\epsilon_{ij}r^j +{a}_i) ,
\ee
where $\rho_e$ stands for electron density and   ${a}_i=(0,
-\fr{emE\te}{2\hbar\ga})$. The expectation value of the current
operator $<{{J}_i}>$ can be calculated with respect to the
eigenstates  of the Hamiltonian (\ref{HG}) given as\cite{ao}
$$
|n,\al,\te> =
\frac{1}{ \sqrt{(2m\hbar \gamma \beta)^n n!} }
\exp\{i(\al y+\frac{m\gamma \beta }{ 2\hbar }xy)\}  (b^{\dagger})^n|0>,
\qquad n=0,1,2...,\qquad \al\in \mathbb{R}.
$$
By definition $b|0>=0.$
Once the current
operator is obtained in terms of creation and annihilation operators,
the calculation is straightforward. Indeed, one can easily show that expectation value of
the $x$-component of current vanishes:
\be <{J}_x> = \langle
n,\al,\te| \fr{e\rho_e\ga}{2 i m}(b-b^\dgr)|n,\al,\te \rangle= 0 .
\ee
On the other hand
expectation value of the $y$--component leads to the Hall conductivity $\si_H :$
\beqa <{J}_y> &=&
\fr{e\rho_e \ga}{m} \langle n,\al,\te |\left( \fr{b+b^\dgr}{2}
-\la_- -\fr{emE\te}{2\hbar\ga } \right)|n,\al,\te\rangle \nno \\
&=& -\fr{e\rho_e }{m}\left( \ga \la_- +\fr{emE\te}{2\hbar } \right)=\si_H  E  .\eeqa

Now we will analyze each Hamiltonian separately:
The coefficients related to (\ref{hr11}), (\ref{hr12}) and (\ref{hr2}) are presented
in Table 1.
\begin{table}[h]
\begin{center}
\begin{tabular}{|c|c|c|c|c|}
\hline \hfill             & $\gamma$    & $\beta$   & $\la_+ $                                     & $\la_- $   \\
\hline $H^{(1)}$   & $  1-2\ka $    & $1$       & $(1-\ka)\la + \fr{emE\te}{ 4\hbar} $ & $(1-\ka)\la - \fr{emE\te}{4\hbar} $ \\
\hline $H^{(2)}$   & $1-2\ka$    & $1-2\ka$  & $(1+2\ka) \la $                              & $\la$ \\
\hline $H^{(3)}$   & $1-4\ka$    & $1$       & $(1+\ka)\la +\fr{emE\te}{4\hbar} $   & $(1+\ka)\la -\fr{emE\te}{4\hbar} $ \\
\hline
\end{tabular}
\end{center}
\caption{Coefficients of the diverse realizations in terms of  $ \ka=\fr{e\te B}{4\hbar c}$ and  $\la=\frac{ m c E}{B} .$}
\end{table}

For the Hamiltonian (\ref{hr12}) the
Hall conductivity does not acquire any $\theta$--deformation:
\be
\lb{hcon2}
 \si_H^{(2)}= -\fr{\rho_e ec}{B} .
\ee
Conversely, one can observe that, although their coefficients differ  the
Hamiltonians (\ref{hr11}) and (\ref{hr2})   give
the same result for the Hall conductivity:
\be
\lb{hcon1}
\si_H^{(1)} (\theta)=\si_H^{(3)}(\theta)  = -\fr{\rho_e e c}{B}(1-\fr{e\te B}{2\hbar c})  .
 \ee
This result is compatible with the one obtained in
\cite{ao} although the deformation factors do not coincide. However, in \cite{ao}
$\theta$--deformation results due to a specific choice of overall factor in the definition of current, here
the current does not possess any unnatural coefficient in its definition.

There are some interesting features. In the
realization (\ref{R1})-(\ref{R2}) if one does not keep the $e^2/c^2$ terms Hall conductivity acquires a deformation factor, in contrary to the
case where the $e^2/c^2$ terms are kept. Higher order corrections in the realization sweeps out the $\theta$ dependence of the Hall
conductivity.  Another curious result is the fact that although their structures are different, two Hamiltonians   (\ref{hr11}) and (\ref{hr2})
lead to the same deformation factor for the Hall conductivity. Consequences of obtaining various types of $\theta$--deformations
of the Hall conductivity will be  discussed in the last section.

\section{Quantum Phases in Noncommutative Space}

The existing formulations of the quantum phases in noncommutative
coordinates can mainly be distinguished by their dependence on
momentum eigenvalues: The formulations of \cite{glr,cpst,mz,wl,prfn} depend on
momentum eigenvalues but the ones in \cite{ao} and \cite{o3} do not
possess any momentum dependence as the original quantum phases.
Except \cite{o3} where a (semi)classical approach was used, all of these formulations implement
noncommutativity in terms of the star product (\ref{NSP}) or the equivalent coordinate shift
(\ref{SH}), however interpretation of the $\theta$ dependent terms appearing in  Hamiltonians differ.  
Because of adopting the   interpretation of \cite{ao}, 
 the $\theta$--deformed phases which we obtain are also  momentum independent.
This issue will be clarified at the end of this section. First, we would like to discuss the existing formulations.
Although in  \cite{glr,cpst,mz,wl,prfn} different phases are considered we will show that they
can be formulated in a unified manner. The starting Hamiltonian operator is
\be
\label{OTH}
H=\frac{1}{2m} \left(  p_\al -\rho A_\al ( r )\right)^2 ,
\ee
where $\rho$ is a constant and the configuration is chosen such that the scalar potential term vanishes. One  implements
noncommutativity by the shift
\be
\label{sfr}
r_\al \rightarrow r_\al -
\frac{1}{2\hbar}\theta_{\al \beta }p^\beta =r_\al -
\frac{1}{2\hbar}\theta_{\al \beta } \left(\hbar  k^\beta +\rho A^\beta (r)\right) ,
\ee
where $k_\al$ is the eigenvalue of the kinetic  momentum
operator:
\be
\label{KMO}
\left( p_\al -\rho A_\al (r)\right)\psi (r)= \hbar k_\al \psi (r).
\ee
Hence, (\ref{OTH}) can be expanded at the first order in $\theta$  as
\be
\label{HOT}
H=\frac{1}{2m} \left[  p_\al -\rho A_\al ( r) + \frac{\rho}{2\hbar}\theta^{\beta\sigma}
\left(\hbar  k_\sigma +\rho A_\sigma ( r)\right)\partial_\beta A_\al ( r) \right]^2.
\ee
Identifying,
\be
\label{aT}
 \tilde{A}_\al ( r, \theta)= A_\al ( r) - \frac{1}{2\hbar}\theta^{\beta\sigma}
 \left(\hbar  k_\sigma +\rho A_\sigma ( r)\right)\partial_\beta A_\al ( r)
\ee
as the gauge field in noncommutative space, one defines the $\theta$--deformed quantum phase by
\be
\label{GDO}
\Phi (\theta ) =\frac{i\rho}{\hbar} \oint  \tilde{A}_\al ( r, \theta) dr^\al .
\ee
Different phases can be considered by choosing the original field $A_\al$ appropriately.
To study the AB phase on the noncommutative plane let the nonvanishing components of the deformation parameter be
$$
\theta_{ij}=\theta\epsilon_{ij},
$$
where $i,j=1,2.$ Moreover, choose $\rho=-e/c$ and
an appropriate 3--vector potential $\bm{A},$ whose third component
vanishes $A_3=0.$ Hence, (\ref{GDO}) leads to
$$
 \Phi_{AB}^I (\theta ) =-\frac{ie}{\hbar c}\oint  \bm{A}( r) \cdot d\bm{r}
-\frac{i m e\theta}{2\hbar c} \oint
\left[\left(\bm{v}\times\bm{\nabla} A_i\right)_3 -\frac{e}{\hbar m
c} \left( \bm{A} \times \bm{\nabla} A_i \right)_3\right] d r_i , 
$$
where $\bm{k}=m\bm{v}.$  This is the deformed phase obtained in
\cite{glr} and \cite{cpst}.

To formulate the
AC, HMW and Anandan phases in noncommutative coordinates we set
\be
\label{gan}
c \rho\bm{A}=
\bm{\mu} \times \bm{E} -\bm{d} \times \bm{B}
 \ee
where   $\bm{\mu}$ and $\bm{d}$ are the magnetic and the electric dipole moments which are proportional to the Pauli spin matrices $\bm \sigma$.
We deal with the standard configuration where dipole moments are in  $z$--direction  and the external electric and magnetic fields are in the polar radial direction,
so that $\bm{\mu}\cdot \bm{B}=0,\ \bm{d}\cdot \bm{E}=0.$ Moreover, let there be no change in the dipoles along
the external fields: $\bm{E}\cdot \bm{\nabla}\mu=0,
\bm{B}\cdot\bm{\nabla}d=0.$ After implying these conditions,
insert (\ref{aT}) into (\ref{GDO})  to obtain
\be \label{GDOA}
\Phi^A (\theta ) =\frac{i}{\hbar c} \oint \left(\bm{\mu} \times \bm{E} -\bm{d}
\times \bm{B} \right) \cdot d\bm{r} +\frac{i}{2\hbar^2c^2} \theta_{ab}\oint
(\bm{k} +\bm{\mu} \times \bm{E} -\bm{d} \times
\bm{B})_a\partial_b (\bm{\mu} \times \bm{E} -\bm{d} \times
\bm{B})\cdot d \bm{r} ,
 \ee
where $a,b=1,2,3.$ For $\bm{d}=0$ the $\theta$--deformation
of the AC phase obtained in \cite{mz} follows
\be
\label{GDO3}
 \Phi^{AC} (\theta ) =\frac{i}{\hbar c} \oint
(\bm{\mu} \times \bm{E} ) \cdot d\bm{r} +\frac{i}{2\hbar^2 c^2}
\theta_{ab}\oint (\bm{k}+\bm{\mu} \times \bm{E} )_a\partial_b
(\bm{\mu} \times \bm{E} )\cdot d \bm{r} .
\ee
For $\bm{\mu}=0,$ the HMW
phase in noncommuting coordinates is obtained
in accord with \cite{wl} as
\be \label{GDO4}
\Phi^{HMW} (\theta ) =-\frac{i}{\hbar c} \oint (\bm{d} \times \bm{B})
\cdot d\bm{r} -\frac{i}{2\hbar^2 c^2} \theta_{ab}\oint (\bm{k}-\bm{d}
\times \bm{B})_a\partial_b (\bm{d} \times \bm{B})\cdot d \bm{r} .
\ee
By putting (\ref{GDO3}) and (\ref{GDO4}) together
$$ 
\Phi^{A1} (\theta )=  \Phi^{AC} (\theta
)+\Phi^{HMW} (\theta ) ,
$$
 which means ignoring the terms behaving as
$\mu d$ in (\ref{GDOA}), the deformation of \cite{prfn}
follows\footnote{There is a discrepancy of sign which seems due to a
misprint in Eq. (33) of \cite{prfn}.}. Although we used
3--dimensional vectors the formalism is effectively 2--dimensional
because of the selected configurations leading to the AC and HMW phases.

The approach of \cite{ao} differs from the above formulation. In \cite{ao} one  considers the $\theta$--deformed Hamiltonian
defined as the generalization of the one obtained in the uniform transverse magnetic field $B.$
In terms of the related path integral one  identifies
$$
\tilde{A}_i(\theta , r) =\left(1- \frac{e\theta B}{4\hbar c}\right)^{-1} A_i (r).
$$
Then, one  employs it in (\ref{GDO}) with $\rho=-e/c$ to get the AB phase in noncommutative coordinates as
$$
\lb{abp} \Phi_{AB}^{II} (\theta ) =-\frac{ie}{\hbar c } \left( 1+ \frac{e\theta B}{4\hbar c}\right)\oint  A_i ( r) dr_i .
$$

Now, let us present our approach following in part the  receipt
given in \cite{ao}. We deal with the configurations leading to
vanishing scalar potentials, so that in general the Hamiltonian in
noncommutative coordinates is written in terms of $\h p$ which is a
realization of the algebra (\ref{rrq})--(\ref{yrq}) as
 \be \lb{HNCD}
 H^{nc}=\frac{ \hat{p}^2}{2m} .
  \ee
Obviously, different realizations
will lead to different Hamiltonians. Let $(r_\al,p_\al)$ define the
classical phase space variables corresponding to the operators
$(r^{op}_\al,\; p_\al^{op}=-i\hbar \partial_\al )  .$ The classical
Hamiltonian $H_{\text{eff}}(r,p)$ will be obtained from the related
Hamiltonian operator in noncommutative space by substituting
$p_\al^{op} , r^{op}_\al $ with the c-number variables $p,r.$ To
keep the discussion general let us define the classical
$\theta$--deformed Hamiltonian corresponding to  (\ref{HNCD}) as \be
\lb{HNC} H^{nc}_{\text{eff}}= a_{\al \beta}(r,\theta) p_\al p_\beta
+ b_\al (r,\theta)p_\al +c (r,\theta), \ee without specifying the
coefficients $a_{\al \beta}(r,\theta),\  b_\al (r,\theta)$ and $c
(r,\theta).$ Plugging  (\ref{HNC}) into  the  path integral \be
\label{pde} Z= N\int d^dp\; d^dr\; \exp \left\{ \fr{i}{\hbar}\int
dt[p^\al\dot{r}_\al -H_{\text{eff}}(p,r)] \right\}, \ee where $N$ is
the normalization factor, yields the partition function in the
$d$--dimensional  phase space:
$$
Z= N\int d^dp\; d^dr\; \exp \left\{\fr{i}{\hbar}
\int dt\, \left[ p^\al \left( \dot{{r}}_\al -  b_\al (r,\al )\right) - a_{\al \beta}(r,\theta) p_\al p_\beta - c (r,\theta) \right]\right\} .
$$
Integration over the momenta
gives the partition function in configuration space with the normalization factor $N^\prime $ as
$$
Z= N^\prime \int d^dr\; \exp \left\{\fr{i}{\hbar} \int
dt \,  \left[ \fr{1}{4} a^{-1}_{\al \beta}(r,\theta)\left( \dot{{r}}_\al
- b_\al (r,\te )\right)\left( \dot{{r}}_\beta -  b_\beta (r,\te
)\right) - c (r,\theta) \right]\right\} .
$$
This  can be written as
$$
Z= N^\prime \int d^dr\; \exp \left\{\fr{i}{\hbar}  S
+\fr{i}{\hbar} \int d{r}_\al     \A^\al (r, \theta ) \right\} ,
$$
in terms of
$$
S = \int \, dt\,
[\fr{1}{4}a_{\al\bt}^{-1}(r, \te)(\dot{r}^\al\dot{r}^\bt+b^\al(r,
\te) b^\bt(r, \te))-c(r, \te)]
$$
and the $\theta$--deformed gauge field defined as
\be
\lb{ABIZ}
\A_\al (r, \theta ) \equiv - \fr{1}{2}a^{-1}_{\al \beta}(r,\theta)  b^\beta (r,\te ).
\ee
Hence, in general we can introduce the quantum phase as follows
\be
\lb{PH}
\Phi= \frac{i}{\hbar} \oint  \A^\al (r, \theta ) dr_\al=
-\frac{i}{2\hbar} \oint a^{-1}_{\al \beta}(r,\theta)  b^\beta (r,\te )dr_\al .
\ee

As the first specific example we would like to discuss the
AB phase in noncommutative space adopting some different realizations.
Hence,
let the  particles be
confined to move on the $r_i=(x,y)$ plane,
 in the presence of an infinitely long,  tiny solenoid placed along the z--axis.
Obviously we set $\rho =-e/c,$ moreover  the nonvanishing components of $\theta$ and $F$ are
$$
\theta_{ij}=\epsilon_{ij}\theta ,\  F_{ij}=\epsilon_{ij}F_{12} =
\left\lbrace
\begin{array}{ll}
\epsilon_{ij} B & {\rm in}, \\
 0 & {\rm out}.
\end{array}
\right\rbrace
$$
Except on the solenoid, the conditions (\ref{cons}) are fulfilled, due to the fact that
$F_{12}$ is constant inside the solenoid and vanishes outside the solenoid.
Thus, we are equipped with the realizations  (\ref{R1})--(\ref{R2}) and (\ref{rea21})--(\ref{rea22})
in a consistent manner. We first deal with the realization given in (\ref{R1}) but ignore the $e^2/c^2$ terms, so that
the related coefficients are
\be
\label{c1}
a_{ij}^{(1)}(r, \te)=\fr{1}{2m} \left(1-\frac{eF_{12}\theta}{2\hbar c}\right)^2 \delta_{ij}, \qquad b_i^{(1)}(r,\te)
=\fr{e}{mc}\left(1-\frac{eF_{12}\theta}{2\hbar c}\right) A_i .
\ee
The trajectory in (\ref{PH}) is chosen to
enclose the origin, thus it yields
\be
\label{NAB1}
 \Phi_{AB}^{nc(1)}=
-\fr{ie}{\hbar c} \oint \left(1+\frac{eF_{12}\theta}{2\hbar c}\right) A_idr_i=
-\fr{ie}{\hbar c} \left(1+\fr{e\te B}{2\hbar c}\right) \int \epsilon_{ij}\nabla_iA_j ds=
\left(1+\fr{e\te B}{2\hbar c}\right) \Phi_{AB} ,
\ee
where the AB phase is given in terms of $\Phi_0=hc/e$ and the cross--sectional area of the solenoid  $S$ as
$$
\Phi_{AB} =-2\pi i\fr{BS}{ \Phi_0} .
$$

When we consider (\ref{R1}) keeping the $e^2/c^2$ terms the coefficients become
\be
\label{NAB2}
a_{ij}^{(2)}(r,
\te)=\fr{1}{2m} \left(1-\frac{eF_{12}\theta}{2\hbar c}\right)^2 \delta_{ij}, \quad b_i^{(2)}(r,
\te)=\fr{e}{mc} \left(1-\frac{eF_{12}\theta}{2\hbar c}\right)^2 A_i.
\ee
Observe that (\ref{ABIZ}) does not acquire any $\theta$--deformation. As a
result of this the phase is not deformed:
\be
\Phi_{AB}^{nc(2)}= \Phi_{AB} \lb{abp2} .
\ee

For the realization  (\ref{rea21}) one can
read the coefficients as follows
\be
\label{NAB3}
a_{ij}^{(3)}(r, \te)=
\fr{1}{2m}\left(1-\frac{eF_{12}\theta}{\hbar c}\right)^2 \delta_{ij},
\quad b_i^{(3)}(r, \te)=-\fr{eB}{2mc} \left(1-\frac{eF_{12}\theta}{\hbar c}\right)\epsilon_{ij}r_j .
\ee
Hence, the $\theta$--deformed AB
phase is deduced as
\be
\lb{abp3}
\Phi_{AB}^{nc(3)}=
\fr{ie}{2\hbar c} \oint  \left(1+\frac{eF_{12}\theta}{\hbar c}\right) F_{12} \epsilon_{ij}r_jdr_i
=\left(1+\fr{e\te B}{\hbar c}\right) \Phi_{AB} ,
\ee
where we used  $S=\oint \epsilon_{ij} r_i dr_j /2.$

Similar to the Hall conductivity (\ref{hcon2}), the realization (\ref{R1}) when the terms at the order of  $e^2/c^2$  are  kept, i.e. (\ref{NAB2}),
does not procure any $\theta$--deformation of the AB phase (\ref{abp2}). However, the other realizations (\ref{c1}) and (\ref{NAB3}) led to
 (\ref{NAB1}) and  (\ref{abp3}) with different $\theta$--dependent factors, in contrary to the Hall effect where they
 yielded the same factor as is given in (\ref{hcon1}). An approach to determine which formulation should be preferred is presented
in the last section.

To discuss the AC, HMW and Anandan phases in noncommutative coordinates we will consider the
realization (\ref{rea21}) in 3 dimensions: $a,b=1,2,3.$ In general it  leads to the
$\theta$--deformed gauge field (\ref{gan}), where
$$
a_{a}^b=\frac{1}{2m}\left(\delta_{a}^b-\frac{2\rho}{\hbar}F_{ac}\theta^{cb}\right)
$$
and
$$
b_a= \frac{\rho}{2m}\left(F_{ab}-\frac{\rho}{\hbar}\theta_{ac}F^{cd}F_{db}\right)r^b.
$$
As far as the conditions (\ref{cons}) are satisfied this construction is valid also for non-Abelian gauge fields.
The $\theta$--deformed
phase factor is
\be
\label{lp}
\Phi^{nc} =
-\fr{i\rho}{2\hbar} \oint \left( F^{ab}+
\fr{\rho}{\hbar}F^{ac}F_{cd}\te^{db} \right)r_a dr_b
\ee
Now we specify the gauge field as in (\ref{gan}) which is appropriate to discuss the AC, HMW and Anandan phases
and consider  the configuration:  $\bm{\mu}=\mu \hat{z},
\bm{d}=d\hat{z};$  $\bm{\mu}\cdot \bm{B}=0,\ \bm{d}\cdot \bm{E}=0$ and $\bm{E}\cdot \bm{\nabla}\mu=0,
\bm{B}\cdot\bm{\nabla}d=0$. Hence the problem is effectively 2--dimensional. The gauge field (\ref{gan}) is now Abelian and the nonvanishing components
of the field strength are
\be
\label{fnv}
 F_{ij}=\epsilon_{ij}(-\mu \bm{\nabla}\cdot
\bm{E}+d\bm{\nabla}\cdot \bm{B}) .
\ee
Moreover, we consider the noncommutative plane by setting
$\theta_{ij}=\epsilon_{ij}\theta .$ As usual
the electromagnetic fields are taken in the radial direction and
their divergence vanish except in  the infinitesimal
regions around the origin where they satisfy\cite{o3}
$$
\bm{\nabla}\cdot\bm{E}=\fr{\la_e}{s^\prm}, \qquad\quad
\bm{\nabla}\cdot\bm{B}=\fr{\la_m}{s^{\prm\prm}}.
$$
We introduced   $s^\prm$ and $s^{\prm\prm}$ which are, respectively, the areas of
the infinitesimal regions where $\bm{\nabla}\cdot\bm{E}$ and $\bm{\nabla}\cdot\bm{B}$ are
nonvanishing. Obviously, $s^\prm$ and $s^{\prm\prm}$  do not play any role in the original definition of
the Anandan  phase:
\be
\label{AO}
\Phi_A=\fr{\rho}{2}\oint F_{ij}r^i dr^j= -\fr{1}{\hbar c}(\mu \la_e-d\la_m) .
\ee
The field strength(\ref{fnv}) satisfies the conditions (\ref{cons}) so that we can use the realization leading to the phase
(\ref{lp}).
The Anandan phase in noncommutative space can be calculated as
\be
\label{nA}
\Phi_A^{nc}= \Phi_A\left[1+\te \left(\fr{\mu\la_e}{\hbar c s^\prm}-\fr{d\la_m}{\hbar c
s^{\prm\prm}}\right)\right] ,
\ee
where  (\ref{AO}) is employed.

Imposing, respectively, $\la_e=0$ and $\la_m=0$  in (\ref{nA}),
the noncommutative AC and HMW phases can be deduced as
\beqa
\Phi_{AC}^{nc} & = & \fr{d\la_m}{\hbar c}\left( 1-\te \fr{d\la_m}{\hbar c s^{\prm\prm}}\right) , \nonumber \\
\Phi_{HMW}^{nc} & = & -\fr{\mu \la_e}{\hbar c}
\left(1+\te \fr{\mu\la_e}{\hbar c s\prm}\right) . \nonumber
\eeqa

Let us compare the constructions of quantum phases in noncommutative coordinates given in
 \cite{glr}--\cite{prfn}  and the ones obtained here. We showed that
the former formulations  result from the $\theta$--deformed gauge field given in (\ref{aT}) which is obtained from the
Hamiltonian (\ref{HOT}) by employing the eigenvalues of the kinetic momentum (\ref{KMO}), so that one gets rid of the momentum operators $p_\al.$
Instead of eliminating  them,
what happens if  one keeps the momentum operators  $p_\al$ in the definition of the $\theta$--deformed Hamiltonian? In this case,
the resulting Hamiltonian would be written as
\be
\label{HOT1}
\tilde{H}=\frac{1}{2m} \left[ \left( \delta_\al^\sigma -\frac{\rho}{2\hbar}\theta^{\beta\sigma}
\partial_\beta A_\al ( r) \right) p_\sigma -\rho A_\al ( r)  \right]^2.
\ee
Now, employing  the corresponding classical Hamiltonian in the path integral (\ref{pde}) and integrating over the momenta would have resulted
in the $\theta$--deformed gauge field as in (\ref{ABIZ}) which is momentum independent.
In fact, the latter procedure is the one which we adopted, though our deformation procedure is different than employing the coordinate  shift
(\ref{sfr}). Hence, interpretation of the $\theta$--deformed terms in Hamiltonians as a contribution to the gauge field or to the kinetic term is the main
difference between these approaches. In our formulation deformed phases are independent of the velocity of the scattered particles
which is one of the main features of the original quantum phases.

\section{Discussions}

We discussed in detail the alternative method of establishing quantum mechanics in noncommutative coordinates which is
applicable to dynamical systems whose  observables are either functions of
phase space variables or matrices which may be independent of phase space coordinates.
The alternative procedure  itself leads to different deformed dynamical systems
depending on the adopted representation of the deformed algebra (\ref{rrq})--(\ref{yrq})  which is equivalent to identify the $\theta$--deformed
quantum phase space variables.

Within  the alternative procedure we discussed the Hall effect in noncommutative coordinates in Section 3 and considered the
quantum phases in noncommutative space in Section 4. Depending on
to the realization adopted the resulting Hall conductivity as well as AB phase
acquire diverse deformation factors in noncommutative coordinates.
Although at first sight this may seem to be a pathological fact, as  we will explain it is  an embarrassment of riches permitting us
to choose the  realization adequate to the problem considered.
One of the interpretations of the noncommutativity of coordinates is
to consider it as an effective method of introducing interactions whose dynamical origins can be complicated\cite{ao,sc}. Once we determine which
realization leads to the desired effective theory we can select to work within that representation. Let us illustrate this considering
the integer quantum Hall effect. Demanding gauge invariance of the extended states
yields quantization of the AB phase gained by the  electrons in Hall effect which permits  one to obtain the integer
quantum Hall effect\cite{LAF}.  
It is possible to extend this formulation to  noncommutative space employing 
noncommutative versions of the Hall effect and the AB phase  obtained in this work.  Depending on 
the  effective theory one desires to obtain by fixing the noncommutativity parameter $\theta$ in the deformed
 integer quantum Hall effect,
 one can select the appropriate deformations of the Hall effect and the AB phase.
Then, the related Hamiltonian can be taken as the as the starting point of formulating a
field theory which may be utilized to find out some other aspects.
For example, we can employ the related noncommutative field theory to
derive Green functions and obtain the quantized Hall effect in noncommutative coordinates similar to the ordinary case\cite{AA}.
These are currently under inspection.

We clarified the relation and discrepancies between our approach and
the existing works on defining the quantum phases in  noncommutative spaces.
We showed that in general quantum phases in noncommutative spaces either for Abelian or for non-Abelian gauge fields
can be defined independent of the velocities of the particles, as the underlying commutative phases.
We discussed the receipt on general grounds and in particularly applied it to the noncommutative plane, due to the fact that
configurations of the observed AB and AC quantum phases are effectively two-dimensional.
Our results can also be applied to condensed matter systems like graphen which are effectively two-dimensional and where the quantum phases play an
important role. This is  an attractive problem because it may give some clues in testing the possible advantages of introducing noncommutative coordinates.

Obviously, we retained the first order contribution in $\theta$ and up to some few orders in coupling constants. However,
our method provides  a systematic receipt of deriving the higher contributions either in $\theta$ or in coupling constants. Moreover, introducing
another deformation parameter similar to $\theta$ to render the  momenta noncommutative even for vanishing $F_{\al\beta},$ is straightforward in our formulation.

\end{document}